# Over 50 mA current in interdigitated diamond field effect transistor

Damien Michez, Juliette Letellier, Imane Hammas, Julien Pernot, and Nicolas Rouger

*Abstract*— This letter presents the bulk diamond field-effect transistor (FET) with the highest current value reported at this moment. The goal was to drastically increase the current of this type of device by increasing the total gate width thanks to an interdigitated architecture and homogeneous growth properties. We report the results obtained by fabricating and characterizing an interdigitated junction FET (JFET). The device develops a total gate width of 14.7 mm, with 24 paralleled fingers and a current higher than 50 mA at $V_{DS}$ = -15 V, $V_{GS}$ = 0 V, at 450 K and under illumination which is the highest value reported for a bulk diamond FET. Its specific ON-resistance and threshold voltage are respectively 608 m$\Omega$.cm², 50 V. From Transfer length method (TLM) measurements we extract a resistivity of 3.6 m$\Omega$.cm for a heavily boron-doped (p++)-diamond layer and 1.52 $\Omega$.cm for a $2.10^{17}$ cm$^{-3}$ p-doped diamond layer at 450 K. We measured the drain current versus gate voltage characteristics at high temperature showing that it is no longer the conduction channel resistance but the device access resistance that is predominant. This study indicates that it is possible to drastically improve the ON-state of FETs by using an interdigitated architecture, while using homogeneous large size diamond layers grown by CVD.

## I. INTRODUCTION

POWER electronics is the key to obtain more sustainable energy management [1]. It is necessary to develop increasingly high-performance transistors for integration into power converters, with low on state resistances and high temperature capability. Wide band gap and ultra-wide bandgap materials have demonstrated efficient and promising devices, such as SiC [2], [3], GaN [4], [5], Ga$_2$O$_3$ [6], [7] and diamond [8], [9], [10], [11]. Diamond's exceptional electrical and thermal properties make it ideal for next-generation power transistors. Diamond has demonstrated a mobility up to 2000 cm².V$^{-1}$.s$^{-1}$ [12], a breakdown field up to 10 MV/cm [13] and a thermal conductivity between 2200 and 2400 W.m$^{-1}$.K$^{-1}$. H-terminated MOSFET showed best performances for the current and the breakdown voltage. However, it is difficult to control the mobility and carrier concentration in the 2DHG (two-dimensional hole gas) channel resistance because of the diamond/oxide interface state [14]. It limits the repeatability of this device type which can be a barrier for industrial development. Bulk conduction FETs (Field Effect Transistor) allow to take advantage of the diamond properties and to have a good doping level, thickness, and so conduction channel resistivity control. Because of the high acceptor activation energy ($\approx 0.38$ eV [12]), boron doped diamond shows a resistance which decreases with the temperature due to the impurity ionization. The charge neutrality equation (1) gives the ionized acceptor density p(T) as a function of the temperature T [12]:

$$p(T) = \frac{1}{2}(\Phi_A - N_D)\left\{\left[1 + \frac{4\Phi_A(N_A - N_D)}{(\Phi_A + N_D)^2}\right]^{1/2} - 1\right\} \quad (1)$$

Where $\Phi_A = \frac{1}{4}N'_V T^{3/2} e^{\frac{-qE_A}{kT}}$ and $N'_V = \frac{2(2\pi m^* k)^{3/2}}{h^3}$

Where k is the Boltzmann constant, h is the Planck constant, q is the elementary charge, m$^*$ = 0.903m$_0$ [15] is the hole effective mass, m$_0$ is the elementary mass of an electron, N$_A$ is the acceptor density, E$_A$ is the acceptor ionization energy and N$_D$ is the donor density (compensation). The low on state resistance at high junction temperatures is unique in bulk diamond devices, and offers a new efficiency/power density trade-off [14]. For power electronics, it is nonetheless necessary to drastically increase previously demonstrated total current capabilities of diamond devices [16], [17], [18], [19], [20], [21]. Hence, interdigitated transistors with large total gate widths are required, above hundreds or thousands of mm, but such large transistors require homogeneous layers with low defect densities. The goal of this work is to increase the total current of bulk diamond FETs by interdigitating and improved diamond layers properties. This study was performed on JFET (Junction FET) but is still valid for other FETs. Fig. 1.a) shows a cross-sectional view of the considered diamond transistor. The boron p-doped layer is used as the conduction channel. The pn junction used as the device gate is formed between this layer and the nitrogen n-doped substrate. Nitrogen has a 1.7 eV [22] activation energy. Consequently, it is necessary to illuminate the device to ionize donors and modulate the junction's space charge region (SCR), thus driving the JFET [23] whenever operated at junction temperature below 200 °C [17].

## II. INTERDIGITATED JFET FABRICATION PROCESS

A (100) HPHT (High Pressure High Temperature) n-doped (nitrogen concentration $\approx 10^{19}$ cm$^{-3}$, thickness $\approx 500$ μm) diamond substrate was used. First, a p-doped layer (target boron concentration = $2.10^{17}$ cm$^{-3}$, thickness = 350 nm) is grown by MPCVD (Microwave Plasma Enhanced Chemical Vapor Deposition). A 1.52 $\Omega$.cm resistivity has been extracted from TLM (Transfer length method) device at 450 K. This layer has been etched by reactive ion etching (RIE) thanks to a O$_2$/CF$_4$ plasma. It allows to mesa-structure the layer to confine the current and electrically insulate the device from each other. A MPCVD heavily boron doped diamond layer (p++) is obtained by selective growth (target boron concentration > $3.10^{20}$ cm$^{-3}$, thickness = 280 nm). A 3.6 m$\Omega$.cm

†This article was submitted for review on April 12, 2024. This work was supported in part by the European project DCADE. This project has received funding from the Clean Sky 2 Joint Undertaking (JU) under grant agreement No 101007868. The JU receives support from the European Union's Horizon 2020 research and innovation program and the Clean Sky 2 JU members other than the Union.

Damien Michez, Julliette Letellier and Imane Hammas are with DIAMFAB, F-38000 Grenoble, France.

Julien Pernot is with the Université Grenoble Alpes, Inst NEEL, CNRS, F-38000 Grenoble, France.

Damien Michez and Nicolas Rouger are with LAPLACE, Université de Toulouse, CNRS, INPT, UPS, Toulouse, France.

resistivity has been extracted from two wires measurements TLM device at 450 K (metal contact surface = 70x50 µm²). Then drain, source and gate metal contacts were deposited by evaporation in two steps, one for the front side (drain and source contacts) and the other for the back side (gate contact).

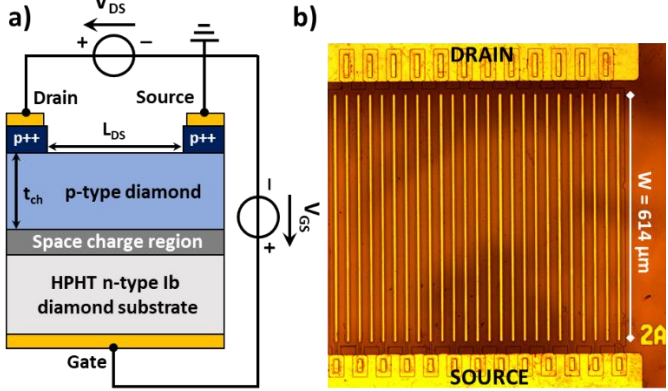

Fig. 1: a) Cross-sectional schematic of an elementary diamond JFET with the electrical measurement setup. b) Optical microscopy top view of interdigitated diamond JFET at the end of the fabrication process.

A Ti/Pt/Au (30/40/50 nm) tri-metal multilayer is used, with titanium, in contact with the diamond to form titanium carbide and have good ohmic contacts on p++ layers. Then an annealing at 600 °C for one hour has been done to favor TiC bonding and reduce the contact resistance [24]. Finally, an ozone treatment has been performed to have a diamond surface oxygen termination and to eliminate surface currents. The ozone plasma was activated thanks to a Xenon EXCIMER UV lamp, centered at 172 nm, and with an oxygen pressure of 500 mbar. An optical microscopy top view of the fabricated interdigitated JFET is shown in Fig. 1.b). The device is composed by 24 transistor fingers, whose elementary cell is described in Fig. 1.a). The drain-source length $L_{DS}$ and the total gate width $W_G$ are respectively 14 µm and 14.7 mm (24x614 µm). This gate width can still be increased in future works, but it is already an important starting point that has never been achieved before on this type of device. Six of these transistors were fabricated on the same sample. The results obtained for each of these transistors showed some disparity, particularly in terms of threshold voltages which are from 50 V to 80 V. This work will focus on the device with the best performance.

### III. INTERDIGITATED JFET CHARACTERIZATION AND RESULTS

As mentioned before, it is necessary to illuminate the JFET to control the gate to source pn junction. All measurements have then been performed under 11 mW/cm² white Leica CLS 150 LED source having a high dark blue component. It should be noted that only photons whose energy is higher than the nitrogen activation (1.7 eV) and lower than the diamond bandgap (5.5 eV) are required to allow the electrostatic control of the pn junction [23]. The electrical measurement setup used in this study consisted of a probe station under vacuum ($\approx 10^{-4}$ mbar), the multisource 2612B Keithley SourceMeter for electrical measurements, the T95-PE LINKAM System Controller coupled with the LNP95 LINKAM Liquid Nitrogen Pump for the temperature control. A Pt100 resistance was used near the diamond die attach to read the actual junction temperature. The pn junction between the sample topside and backside was measured at 300 K and under illumination. The normalized forward resistance is 31 MΩ.cm² (interdigitated JFET pn junction surface = 630x727 µm²). The OFF state was measured at 230 K to reduce leakage current and on a non-interdigitated JFET to protect interdigitated JFETs and so to continue their study (Fig. 2.a).

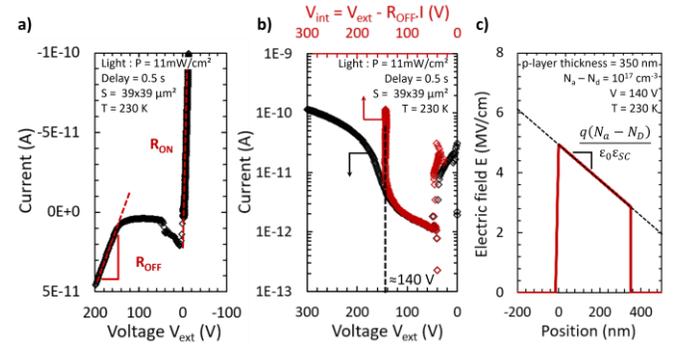

Fig. 2: Non-interdigitated JFET pn junction characterization. a) I-V ON state and OFF state characteristics. b) Estimation of the volage before leakage current emergence. c) Electric field 1D simulation in the pn junction SCR's. $\varepsilon_0$ is the vacuum permittivity constant and $\varepsilon_{SC}$ is the diamond relative dielectric constant.

In Fig 2. b), the pn junction current has been plotted in red as a function of $V_{int} = V_{ext} – R_{OFF}·I$ where $V_{int}$ is the internal pn junction potential and $V_{ext}$ is the measured voltage (here the OFF-state resistance $R_{OFF}$ = 75 MΩ.cm²). $R_{OFF}$ is different from $R_{ON}$ (1.5 MΩ.cm²) the resistance in the ON state (Fig. 2.a). It shows that the substrate resistance does not limit the current in the OFF state. On the I-$V_{int}$ curve there is no significant leakage current before $V_{ext}$ = 140 V. At this point $R_{OFF}·I$ is no longer negligible and $V_{int}$ becomes constant. It corresponds to a 5 MV/cm maximum electric field in the junction for the 1D simulation in the punch-trough configuration (Fig. 2.c). The drain current ($I_D$) versus drain voltage ($V_{DS}$) characteristic $I_D$-$V_{DS}$ was measured (Fig. 3.a). The gate voltage $V_{GS}$ was swept between -10 V and 50 V with 10 V steps. The current shown by the interdigitated JFET is over 50 mA ($\approx$ 3.5 mA/mm) at $V_{DS}$ = - 15 V and $V_{GS}$ = 0 V and 450 K which is the highest value reported for a bulk diamond FET (Fig. 4.a). From the Fig. 3.b), the measured threshold voltage is around 50 V. Measurements were done with a 0.1 s delay, defined as the voltage bias duration before the current is measured. This delay is a trade-off between overall measurement speed and reaching a steady state condition on the internal bias of the pn junction. This value corresponds to a gate voltage sweep of 10 V/s. The hysteresis between up and down gate voltage sweeps comes from the substrate high resistivity. It significantly increases the time constant τ = R·$C_{iss}$ of the RC circuit formed by the substrate and the pn junction's SCR (with the input capacitor $C_{iss}$ = $\varepsilon_0·\varepsilon_{SC}·S/W_{SCR}$, where $W_{SCR}$ is the depleted region thickness, $\varepsilon_0$ is the vacuum permittivity constant and $\varepsilon_{SC}$ is the diamond relative dielectric constant). $C_{iss}$ is defined as $C_{iss}$ = $C_{GS}$ + $C_{GD}$ ($C_{GS}$ and $C_{GD}$ are the gate-source and the gate-drain capacitors) and its surface S is the total device surface (727x630 µm²). For $V_{GS}$ = $V_{GD}$, the time constant can be estimated by considering the SCR between 120 nm ($V_{GS}$ = 0 V) and 350 nm (fully depleted, $V_{GS}$ > $V_{th}$). In these cases, $C_{iss}$ is ranging between 186 pF and 64 pF, corresponding to τ ranging between 1.24 s and 0.43 s, in good agreement with the observed hysteresis and result in [23], for the same device type. The wavelength dependence of these parameters has been studied in [25].

In the OFF state at $V_{DS}$ = -1 V and $V_{GS}$ > 60 V, the gate leakage current is twice the drain current because it is shared between the drain and the source electrodes. It shows that there is no any other leakage mechanism between drain and source, under these conditions. In the ON state, for $V_{DS}$ = -1 V and $V_{GS}$ < -10 V, the gate leakage current is almost six orders of magnitude lower than the drain to source current. The Fig. 4.a) represents the ON state current for a |$V_{DS}$| = 15 V absolute voltage as a function of the specific resistance $R_{ON}$.S in the state of the art. The device design has not been fully optimized, and it shows a 600 mΩ.cm² specific on state resistance at 450 K. It was calculated by taking the measured ON-state resistance $R_{ON}$ at $V_{GS} = 0$V (Fig. 3.a) normalized by the total conduction channel surface $W_G \cdot L_{DS}$.

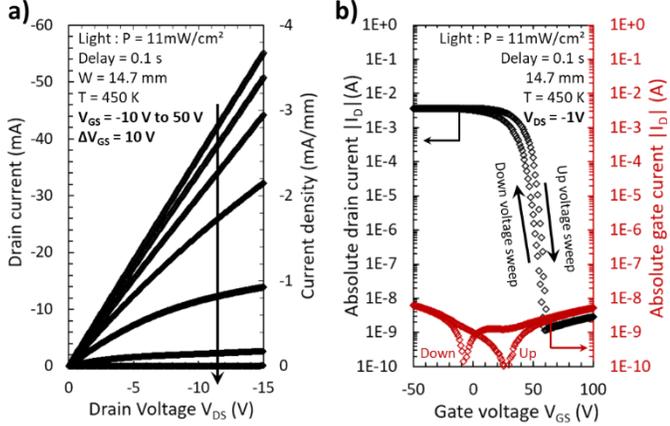

Fig. 3: Interdigitated JFET's a) Drain current ($I_D$) versus drain voltage ($V_{DS}$) characteristic $I_D$-$V_{DS}$. $V_{GS}$ is varied from -10 V to 50 V with 4 V steps. b) Drain current ($I_D$) versus gate voltage ($V_{GS}$) characteristic $I_D$-$V_{GS}$.

In the state of the art shown in Fig. 4a , this specific resistance is not the lowest: for example [18] is better (lower Ron.S), considering its conduction channel surface $W_G \cdot L_{GD}$. The devices presented in the state of the art were not optimized for a large total current capability; a performance they could have but was not experimentally demonstrated in the prior art. In this work, the goal was to drastically increase the total current through a larger total gate width (interdigitated architecture) and homogeneous large size diamond layers grown by CVD. In this way, the largest total current value for a bulk diamond FET is achieved for the interdigitated JFET presented in this work.

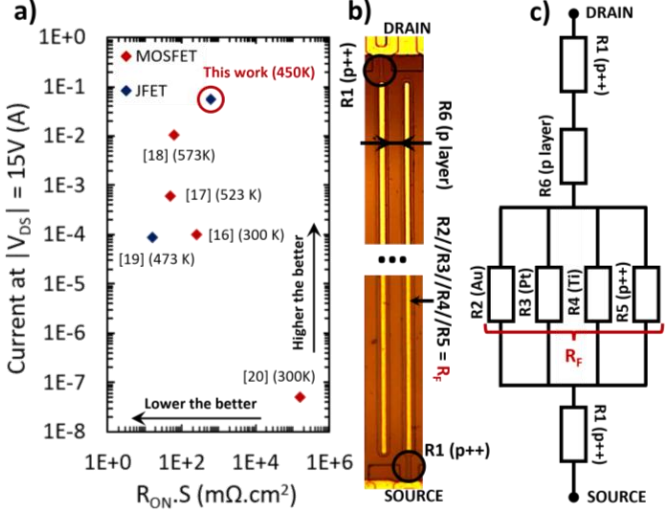

Fig. 4: a) ON state current for a $|V_{DS}|= 15$ V absolute voltage as a function of the specific resistance $R_{ON}.S$ state of the art [17], [18], [19], [20]. b) Zoom on one finger of interdigitated diamond JFET optical microscopy top view. Resistances between drain and source are represented. c) Equivalent diagram of resistors limiting the current between drain and source.

Fig. 4.b) and c) show that $R_{ON}$ is composed of several resistances in series. The first, R1, is only constituted by the p++ layer bottleneck (S = 5x32 µm²). Then there is the finger resistance $R_F$ which corresponds to metal layers (S = 5x614 µm²) and p++ layer (S = 15x614 µm²) in parallel (R2//R3//R4//R5). Both previous resistances composed what we consider to be the access resistance $R_A$ of the device ($R_A$ = 2R1+$R_F$). The last resistance R6 corresponds to the conduction channel (S = 14x614 µm²). The total resistance $R_{ON}$ = ($R_A$+R6)/24 due to the 24 transistors in parallel. This model was used to fit the device resistance as function of the temperature.

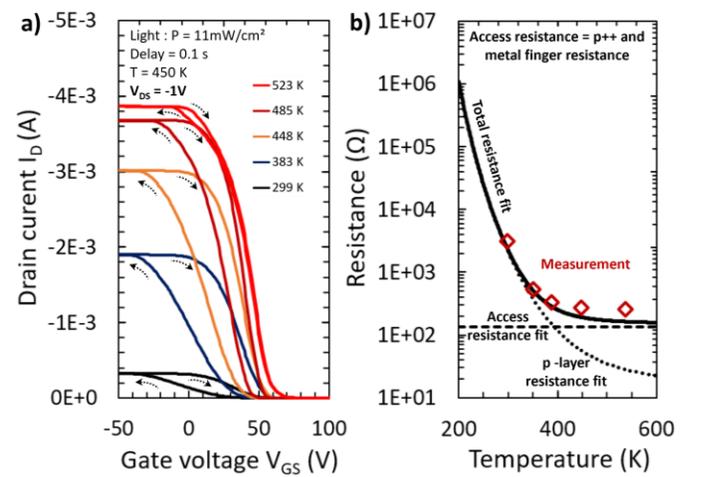

Fig. 5: a) $I_D$ -$V_{GS}$ characteristics at 299 K, 383 K, 448 K, 485 K and 523 K. b) Interdigitated JFET's resistance as a function of the temperature. Measurements are in red, and simulations in black.

The Fig. 5.a) shows the $I_D$-$V_{GS}$ characteristics at room temperature (299 K), 383 K, 448 K, 485 K and 523 K. It provides an overview of the JFET performances regarding the temperature (threshold voltage, total resistance of the linear regime, drain current and hysteresis). The measurements show the linear regime, the transconductance and pinch off regime. The hysteresis decreases with increasing temperatures. As explained before, this hysteresis probably comes from the substrate resistance which is nitrogen doped. Donor ionization follows the same model as acceptor ionization. By increasing the temperature, more donors are ionized and the gate internal resistance decreases. It explains why the voltage amplitude of the hysteresis decreases with the temperature. The simulated total transistor resistance as a function of temperature has been plotted in black (Fig. 5.b) from the model described in Fig. 4.b) and c), taking into account all layer contributions. Red points are measured resistances from the Fig. 5.a). At low and high temperatures, the conduction channel and access resistance respectively limit the current. Here, the 3.6 mΩ.cm TLM extracted resistivity (p++ layer) is considered but these devices are near the edge of the sample and the transistor is near the center. It corresponds to a doping level of 6 x $10^{20}$ cm$^{-3}$ (typical mobility ≈ 3 cm²V$^{-1}$s$^{-1}$ [26]). Differences between measurements and fits probably comes from p++-layer doping level homogeneity. Nevertheless, the resistance model used is a good description of the device presented in this work.

## IV. CONCLUSION

A bulk diamond interdigitated JFET has been demonstrated. Its total gate width is 14.7 mm. The threshold voltage $V_{th}$ and the specific resistance $R_{ON}.S$ are respectively 50 V and 600 mΩ.cm². It showed a current higher than 50 mA at $V_{DS}$ = - 15 V, $V_{GS}$ = 0 V and 450 K which is the highest value reported for a bulk diamond FET. The resistance model shows that the access resistance limits the current at high temperature which can be changed by removing the bottleneck p++-doped layer. The presented device can be optimized, but this work shows that the diamond is a sufficiently mature material to fabricate interdigitated devices as required by power electronics.

## ACKNOWLEDGEMENT

The authors would like to thank NANOFAB and SC2G team from the Néel institute for device fabrication and characterization support.